\documentclass[]{aa}  
\usepackage{graphicx}
\usepackage[varg]{txfonts}

\usepackage{epstopdf,color}
\begin{document}

   \title{Near-infrared photometry of WISE~J085510.74$-$071442.5}

%  \subtitle{I. Overviewing the $\kappa$-mechanism}

   \author{M. R. Zapatero Osorio
          \inst{1}
          \and
          N. Lodieu\inst{2,3}%\fnmsep\thanks{Just to show the usage of the elements in the author field}
          \and
          V. J. S. B\'ejar\inst{2,3}
          \and
          E. L. Mart\'in\inst{1}
          \and
          V. D. Ivanov\inst{4}
          \and
          A. Bayo\inst{5}
          \and
          H. M. J. Boffin\inst{4}
          \and
          K. Mu\v{z}i\'c\inst{6}
          \and
          D. Minniti\inst{7,8,9}
          \and
          J. C. Beam\'in\inst{5,8}
          }

   \institute{Centro de Astrobiolog\'ia (CSIC-INTA), Carretera de Ajalvir km 4, E-28850 Torrej\'on de Ardoz, Madrid, Spain.\\
               \email{mosorio@cab.inta-csic.es}
          \and
              Instituto de Astrof\'isica de Canarias, C/. V\'ia L\'actea s/n, E-38205 La Laguna, Tenerife, Spain.
 %            \email{nlodieu@iac.es,vbejar@iac.es}
 %            \thanks{The university of heaven temporarily does not accept e-mails}
          \and
              Departamento de Astrof\'isica, Universidad de La Laguna, E-38206 La Laguna, Tenerife, Spain.
          \and
              European Southern Observatory, Alonso de C\'ordova 3107, Vitacura, Santiago, Chile.
          \and
              Instituto de F\'isica y Astronom\'ia, Facultad de Ciencias, Univ$.$ de Valpara\'iso, Av. Gran Breta\~na 1111, Valpara\'iso, Chile.
          \and
              N\'ucleo de Astronom\'ia, Facultad de Ingenier\'ia, Univ. Diego Portales, Av. Ejercito 441, Santiago, Chile.
          \and
              Dpt$.$ Ciencias F\'isicas, Univ$.$ Andr\'es Bello, Campus La Casona, Fern\'andez Concha 700, Santiago, Chile.
          \and
              The Millennium Institute of Astrophysics, Santiago, Chile.
          \and
              Vatican Observatory, V00120 Vatican City State, Italy.
             }

   \date{Received ; accepted }

% \abstract{}{}{}{}{} 
% 5 {} token are mandatory
 
  \abstract
  % context heading (optional)
  % {} leave it empty if necessary  
   {}
  % aims heading (mandatory)
   {We aim at measuring the near-infrared photometry, and deriving the mass, age, temperature, and surface gravity of WISE~J085510.74$-$071442.5 (J0855$-$0714), which is the coolest known object beyond the Solar System as of today. }
  % methods heading (mandatory)
   {We use publicly available data from the archives of the {\sl Hubble Space Telescope} ({\sl HST}) and the Very Large Telescope (VLT) to determine the emission of this source at 1.153 $\mu$m ($F110W$) and 1.575 $\mu$m ($CH_4$-off). J0855$-$0714 is detected at both wavelengths with signal-to-noise ratio of $\approx$10 ($F110W$) and $\approx$4 ($CH_4$-off) at the peak of the corresponding point-spread-functions.  }
  % results heading (mandatory)
   {This is the first detection of J0855$-$0714 in the $H$-band wavelengths. We measure the following magnitudes: 26.31\,$\pm$\,0.10 and 23.22\,$\pm$\,0.35 mag in $F110W$ and $CH_4$-off (Vega system). J0855$-$0714 remains unresolved in the {\sl HST} images that have a spatial resolution of 0.22\arcsec. Companions at separations of 0.5 AU (similar mass and brightness) and at $\sim$1 AU ($\approx$1 mag fainter in the $F110W$ filter) are discarded. By combining the new data with published photometry, including non-detections, we build the spectral energy distribution of J0855$-$0714 from 0.89 through 22.09 $\mu$m, and contrast it against state-of-the-art solar-metallicity models of planetary atmospheres. We determine that the best spectral fit yields a temperature of 225--250\,K, a bolometric luminosity of log\,$L/L_\odot$ = $-$8.57, and a high surface gravity of log\,$g$\,=\,5.0 (cm\,s$^{-2}$), which suggests an old age although such a high gravity is not fully compatible with evolutionary models. After comparison with the cooling theory for brown dwarfs and planets, we infer a mass in the interval 2--10 M$_{\rm Jup}$ for ages of 1--12 Gyr and high  atmospheric gravities of log\,$g \gtrapprox 3.5$ (cm\,s$^{-2}$). If it has the age of the Sun, J0855$-$0714 would be a $\approx$5-M$_{\rm Jup}$ free-floating planetary-mass object.}
  % conclusions heading (optional), leave it empty if necessary 
   {J0855$-$0714 may represent the old image of the free-floating planetary-mass objects of similar mass discovered in star-forming regions and young stellar clusters. From the extrapolation of the substellar mass functions of young clusters to the field, as many J0855$-$0714-like objects as M5--L2 stars may be expected to populate the solar neighborhood. }

   \keywords{planetary systems -- brown dwarfs -- stars: low-mass -- stars: late-type 
            }

   \maketitle
%
%-------------------------------------------------------------------

\section{Introduction \label{intro}}
The existence of free-floating, planetary-mass objects with masses near and below the deuterium-burning limit at 13 Jupiter masses (M$_{\rm Jup}$) and temperatures below $\sim$2200 K is established in several star-forming regions and open clusters younger than $\sim$150 Myr, including Orion \citep{lucas00,lucas01,osorio00,barrado01,weights09,bayo11}, $\rho$\,Ophiucus \citep{marsh10}, Upper Scorpius \citep{lodieu13}, Chamaeleon-I \citep[and references therein]{muzic15}, and the Pleiades \citep{osorio14a,osorio14b}. At the age of a few Gyr, substellar evolutionary models predict extremely low temperatures (typically $\le$500 K), similar to the planets of our Solar System, for these isolated planetary-mass objects \citep{chabrier00,burrows03}. To date, there is only one unique known object of this kind: WISE\,J085510.83$-$071442.5 discovered by \citet[][hereafter J0855$-$0714]{luhman14a}; it likely represents the cold and old version of the young, free-floating planetary-mass objects.

J0855$-$0714, found after a careful analysis of multi-epoch astrometry from the {\sl Wide-field Infrared Survey Explorer} ({\sl WISE}) and the {\sl Spitzer Space Telescope} data, has a high proper motion of 8.13\,$\pm$\,0.3 arcsec\,yr$^{-1}$ and is located at a distance of 2.31\,$\pm$\,0.08 pc \citep{luhman14b,wright14}, which makes it the fourth closest known system to the Sun. The most striking property of J0855$-$0714 is its extremely cool nature as inferred from the very red colors $[3.6]-[4.5]$, $W1-W2$, and $J-[4.5]$ \citep{faherty14}. \citet{luhman14a} estimated a temperature of $T_{\rm eff}$\,=\,225--260 K, thus confirming J0855$-$0714 as the coldest substellar object found in isolation with a temperature that fills the gap between transiting exoplanets and our Solar System planets. Several groups have attempted to detect J0855$-$0714 at optical and near-infrared wavelengths using ground-based facilities but with little success. \citet{kopytova14}, \citet{beamin14}, and \citet{wright14} reported upper limits of 24.8 mag, 24.4 mag, and 22.7 mag in the $z$-, $Y$-, and $H$-bands, respectively, while the $K$-band upper limit of 18.6 mag comes from the Visible and Infrared Survey Telescope for Astronomy \citep[VISTA,][]{emerson06} Hemisphere Survey \citep{mcmahon13} using the VISTA Infrared Camera \citep[VIRCAM,][]{dalton06}. The only ground-based reported tentative detection is in the $J$-band ($J$\,=\,25.0\,$\pm$\,0.5 mag) although with a modest signal-to-noise ratio of 2.6\,$\sigma$ \citep{faherty14}. With all these photometric measurements in hand, current models predict temperatures well below $\sim$500K and masses in the planetary domain for J0855$-$0714 (see \citealt{leggett15}). As a reference, the temperature of Jupiter's atmosphere at 1 bar is 165 K.

Here, we report the first clear near-infrared detections of J0855$-$0714 at 1.153 and 1.575 $\mu$m using public archival data from the European Southern Observatory (ESO) and the {\sl Hubble Space Telescope} ({\sl HST}). We employed these data together with published photometry to improve the derivation of the spectral energy distribution and to constrain the properties of this intriguing planetary-mass object. The same data are also presented in \citet{luhman16}, where these authors discuss the new near-infrared photometry of J0855$-$0714 in comparison with other Y dwarfs and model atmospheres.

\section{Observations \label{obs}}

\begin{figure*}
%\resizebox{\hsize}{!}
\center{\includegraphics[scale=0.14]{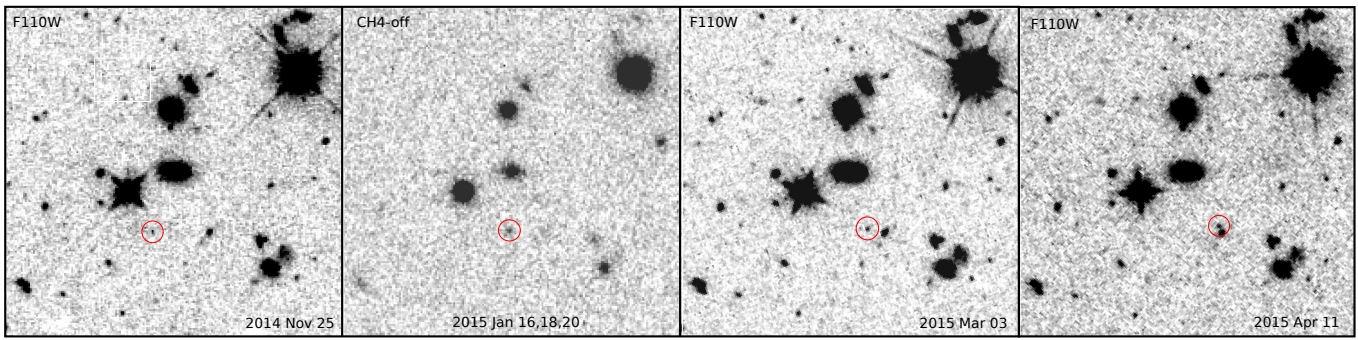}}
\caption{Identification of WISE~J085510.74$-$071442.5 (at the center of the red circles) on the WFC3 and Hawk-I images sized 24\arcsec\,$\times$\,24\arcsec. The observing dates and filters are indicated. North is up and East is to the left.
      \label{fc}}
\end{figure*}

J0855$-$0714 was observed with the $F110W$ filter (centered at 1.1534 $\mu$m, passband of 0.5 $\mu$m) and the Wide-Field Camera 3 (WFC3) on-board the {\sl HST} on three different occasions (2014 Nov 25, 2015 Mar 03, and 2015 Apr 11) under program number 13802 (PI: K$.$ Luhman). We downloaded the reduced WFC3 frames from the Mikulski Archive for Space Telescopes\footnote{https://archive.stsci.edu/hst/}, which include flux calibrated, geometrically-corrected, and dither-combined images processed with the CALWF3 code version 3.3. The total exposure time was 5417.6 s per observing epoch. The released images have a plate scale of 0.1285\arcsec\,pixel$^{-1}$ and a spatial resolution of 0.22\arcsec~as measured from the full-width-at-half-maximum (fwhm). The three public epochs are separated by 98 d (first and second) and 38.3 d (second and third), and they have a field of view of approximately 2\arcmin$\times$2\arcmin; because of its high motion and significant parallax, J0855$-$0714 displaced itself by $\sim$20 and $\sim$10 WFC3 pixels between the first and second, and the second and third {\sl HST} images, respectively. J0855$-$0714 is detected in the broad-band $F110W$ filter on the three occasions with a signal-to-noise ratio of about 10 at the peak flux. Figure~\ref{fc} shows a portion of the WFC3 images with the identification of our target. Using the {\em phot} package within IRAF\footnote{The Image Reduction and Analysis Facility (IRAF) is distributed by National Optical Astronomy Observatories, whcih is operated by the Association of Universities for Research in Astronomy, Inc., under contract with the National Science Foundation.}, we obtained the $F110W$ photometry for an aperture of 6 pixels (or 0.8\arcsec), a sky annulus between 8 and 15 pixels, and a zeropoint of 26.0628 mag (as of the instrument calibration\footnote{http://www.stsci.edu/hst/wfc3/phot\_zp\_lbn} of 2012 Mar 06, Vega photometric system). The photometry of the third {\sl HST} epoch was obtained with a small aperture radius of 2 WFC3 pixels because J0855$-$0714 lies very close to another source (see Figure~\ref{fc}). An aperture correction was later applied to bring the derived instrumental photometry to the apparent value used in this paper. The WFC3 frames are astrometrically calibrated; the coordinates of J0855$-$0714 were determined from the centroids given by {\em phot}. The derived magnitudes and the astrometry of J0855$-$0714 as a function of observing date are given in Table~\ref{phot}.

\begin{table}
\caption{\label{phot}New photometry and astrometry of J0855$-$0714.}
\centering
\tiny
\tabcolsep=0.11cm
\begin{tabular}{llccl}
\hline\hline
Instrument & MJD & RA (J2000) & Dec (J2000) & Photometry \\
           &     & (h m s) & ($^{\rm o}$ $'$ $''$) & (mag) \\
\hline
%WFC3   & 56986.778 & 8 55 08.433 & $-$7 14 39.49 & $F110W$\,=\,26.36\,$\pm$\,0.15  \\
%HAWK-I & 57040.165\tablefootmark{a} &       &       & $CH_4$\,=\,23.22\,$\pm$\,0.35 \\
%WFC3   & 57084.752 & 8 55 08.248 & $-$7 14 39.29 & $F110W$\,=\,26.31\,$\pm$\,0.10  \\
WFC3   & 56986.817 & 8 55 08.433 & $-$7 14 39.49 & $F110W$\,=\,26.36\,$\pm$\,0.15  \\
HAWK-I & 57040.165\tablefootmark{a} &       &       & $CH_4$\,=\,23.22\,$\pm$\,0.35 \\
WFC3   & 57084.818 & 8 55 08.248 & $-$7 14 39.29 & $F110W$\,=\,26.31\,$\pm$\,0.10  \\
WFC3   & 57123.166 & 8 55 08.163 & $-$7 14 39.34 & $F110W$\,=\,26.32\,$\pm$\,0.10  \\
\hline
\end{tabular}
\tablefoot{Right ascension (RA) and declination (Dec) coordinates are given with a precision of $\pm$0.05\arcsec. Photometry is in the Vega system.\\
\tablefoottext{a}{Only the modified Julian date corresponding to the 2015 Jan 18 epoch is given. These data are not astrometrically calibrated. }
}
\end{table}

J0855$-$0714 was also observed with the methane filter ($CH_{4}$-off, centered at 1.575 $\mu$m, passband of 0.112 $\mu$m) of the High Acuity Wide field $K$-band Imager \citep[HAWK-I;][]{pirard04,casali06,kissler_patig08,siebenmorgen11} mounted on the Nasmyth A focus of the European Southern Observatory Very Large Telescope (VLT) unit 4. The on-sky field of view is 56.25 arcmin$^2$ with a cross-shaped gap of 15\arcsec~between the four HAWAII 2RG 2048$\times$2048 pixels detectors. The expected location of J0855$-$0714 lies on the north-west detector. The pixel scale is 0.106\arcsec. We downloaded the public raw data obtained on different occasions as part of program 094.C-0048 (PI: K$.$ Luhman) and reduced the data for each date separately using the {\tt{esorex}} pipeline (version 3.12). The tasks performed by the HAWK-I pipeline included the creation of a master dark and master twilight flat-field as well as the reduction of the jitter observations up to the final reduction stage, which incorporated the alignement and combination of sky-subtracted individual images. We did not run the recipe dealing with the zeropoint magnitude because it is not available for the methane filter. Of all images publicly available, only three epochs provided the deepest data, which we use here: 2015 Jan 16, 18 and 20. The seeing was 0.4\arcsec--0.5\arcsec, and on-source exposure times were 2500 s (Jan 16) and 5000 s (Jan 18 and 20). These observing dates are bracketed by the {\sl HST} ones, which allowed us to predict the position of J0855$-$0714 with a high accuracy. Our target is detected with a weak signal at the expected location in each individual date. Over the period of 4 days, J0855$-$0714 moves less than one HAWK-I pixel. Therefore, to improve the quality of the detection without degrading the spatial resolution of the original data, we combined the three images into one, a portion of which is illustrated in Figure~\ref{fc}. J0855$-$0714 is unambiguously seen with a signal-to-noise ratio of 4 at the peak flux. This is the first time that J0855$-$0714 is detected at the $H$-band wavelengths from the ground. We performed the photometric calibration of the methane images by adopting a null (neutral) $H-CH_{4}$-off color for three 2MASS stars \citep{skrutskie06} that are not saturated in the field covered by the fourth detector. We then obtained the point-spread-function photometry of J0855$-$0714 deriving $CH_4$-off\,=\,23.22\,$\pm$\,0.35 mag (Table~\ref{phot}), where the error bar accounts for the photon noise of the target and the uncertainty of the photometric calibration. The obtained $F110W$ and $CH_4$-off photometry is compatible with the $F125W$ and $F160W$ data recently reported by \citet{schneider16}. The {\sl HST} and VLT observing journal is provided in Table~\ref{log} of the Appendix.

\section{Variability, astrometry and search for companions}
J0855$-$0714 does not show evidence of photometric variability with amplitudes larger than $\sim$0.1 mag at $F110W$. However, we caution that the small difference between the three {\sl HST} detections may not mean low amplitude of variability. Very low-mass dwarfs are known to be fast rotators at nearly all ages. For example, the $\sim$10-Myr planet 2M1207b \citep{chauvin05} rotates with a period of 10.7\,$^{+1.2}_{-0.6}$ h \citep{zhou16}, and the two older brown dwarf components of the Luhman\,16AB system (the closest known brown dwarfs, \citealt{luhman13}) rotate with a period of 5.1 $\pm$ 0.1 h (the B component) and 4--8 h (A) \citep{burgasser14,buenzli15,mancini15}. As a reference, Jupiter has a sidereal rotation period of 9.925 h\footnote{Jupiter fact sheet: http://nssdc.gsfc.nasa.gov/planetary/factsheet/jupiterfact.html}. Even faster rotations of $\approx$2 h have been reported in the literature for several brown dwarfs \citep[e.g.,][]{clarke02,williams15}. The {\sl HST} observations of J0855$-$0714 cover 2.35, 3.72 and 5.17 h on the three observing epochs (Table~\ref{log}). If the rotation of our target is of the order of hours, the $F110W$ data would have averaged the object's flux over a significant fraction of the rotational period, which could smooth the variability to a small magnitude difference.

Using the published astrometry of J0855$-$0714 \citep{luhman14b} and the new measurements from the WFC3 data (Table~\ref{phot}), we determined a new parallax following the procedure described in \citet{osorio14c}. The values obtained, including nine epochs of observations between 2010.34 and 2015.28, are the following: $\mu_\alpha$\,=\,$-8.16 \pm 0.05$ arcsec\,yr$^{-1}$, $\mu_\delta$\,=\,$+0.66 \pm 0.05$ arcsec\,yr$^{-1}$, $\pi$\,=\,$0.464 \pm 0.020$ arcsec, which translates into a distance of $d$\,=\,2.16\,$\pm$\,0.10 pc. These measurements are consistent within 1\,$\sigma$ the quoted uncertainties with those of \citet{luhman14b}, thus confirming that the distance to J0855$-$0714 is solidly established.

The excellent spatial resolution of the WFC3 images allowed us to constrain the multiplicity nature of J0855$-$0714 at separations $\ge$0.5 AU (provided the trigonometric distance of 2.2 pc). We investigated the presence of any co-moving object within a radius of 50 AU (or $\approx$22\arcsec). At the shortest separations of 0.5 AU, J0855$-$0714 appears unresolved; therefore, companions of similar brightness (or mass) are discarded. At distances of $\ge$1 AU from the central object, no other source shows a high proper motion comparable to that of our target; hence, companions with $F110W$ brightness up to $\approx$1 mag fainter (4 $\sigma$) than J0855$-$0714 can also be ruled out. If J0855$-$0714 has any companion, it would lie at a projected orbit of semi-major axis likely less than 0.5 AU. Very accurate astrometry may reveal the presence of disturbances in the coordinates of J0855$-$0714 that could be due to close companions (other planet-hunter techniques, like radial velocity studies, are not applicable to J0855$-$0714 because of its intrinsically faint luminosity and the lack of stable, high-resolution spectrographs operating at mid-infrared wavelengths). From the parallax and proper motion solution, we obtained astrometric residuals (i.e., observed minus computed values) that are typically within 3-$\sigma$ the quoted astrometric uncertainties. A more precise astrometric study can be carried out by considering the relative phase artificially introduced by the location of the different space-based observatories. We did not account for this effect here.

\section{Temperature and gravity \label{fit}}
\begin{figure}
%\resizebox{\hsize}{!}
\center{\includegraphics[scale=0.48]{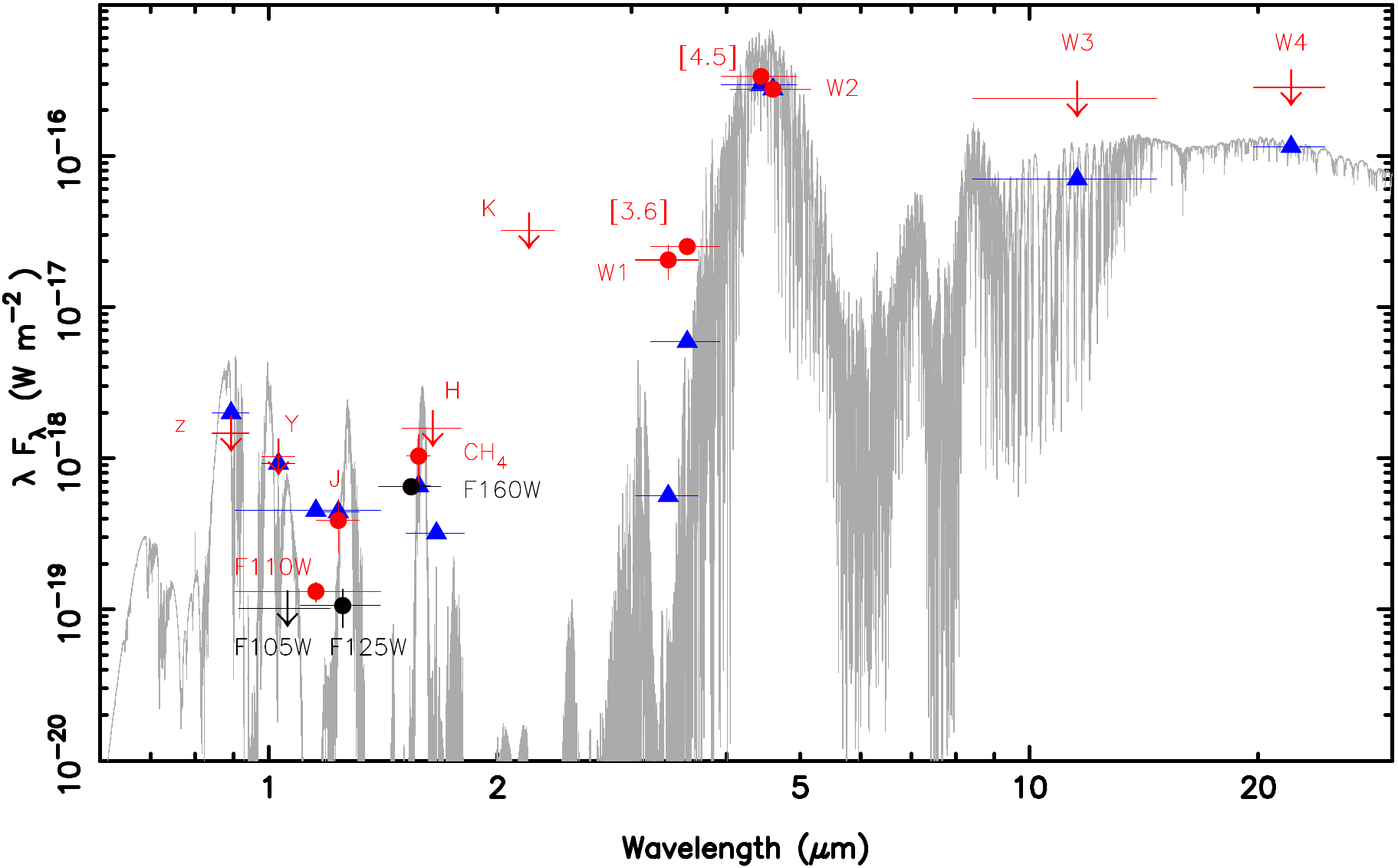}}
\caption{Spectral energy distribution of J0855$-$0714 (red and black symbols). Circles denote positive detections and arrows indicate upper limits. The black symbols correspond to \citet{schneider16} data. The horizontal error bars account for the width of the various filters. The filters are labeled. The best fit planetary model atmosphere of \citet{morley14} computed as the average of $T_{\rm eff}$\,=\,225 and 250 K, log\,$g$\,=\,5.0 (cm\,s$^{-2}$) and 50\%~cloudy conditions is also shown with a gray line. The blue triangles represent the theoretical flux densities as integrated from the models using the corresponding filter passbands (only for the filters in red color). The new detections of J0855$-$0714 presented here correspond to $F110W$ and $CH_4$. The model is normalized to the emission of J0855$-$0714 at the wavelengths of the $W2$ filter.
      \label{sed}}
\end{figure}

We built the photometric spectral energy distribution (SED) of J0855$-$0714 by converting our photometry and the photometry available in the literature (see Section~\ref{intro}) into observed flux densities. We used the Vega flux densities of 1784.9 Jy \citep{schultz05} and 1048.801 Jy \citep{cohen92} at $F110W$ and $CH_4$-off, respectively. For the remaining filters we employed the flux densities given in \citet{reach05} for {\sl Spitzer}, \citet{hewett06} for $z$, $Y$, $J$, $K$, and \citet{jarrett11} for {\sl WISE}. The resulting SED is shown in Figure~\ref{sed}, where clear detections are plotted with a solid symbol and arrows indicate upper limits on the fluxes imposed by limiting magnitudes quoted in the literature. For completeness, we also included the {\sl HST} photometry of \citet{schneider16} in Figure~\ref{sed}. Even non-detections are relevant to study the SED of J0855$-$0714. The emission of this object is highest at $\sim$4.5 $\mu$m and shows a sharp increase by about three orders of magnitude from the near-infrared wavelengths to the peak of the SED. In the near-infrared, the largest signal occurs at the $CH_4$-off filter because the narrow width of this passband avoids the part of $H$-band strongly absorbed by methane and only registers frequencies less affected by water vapor and methane absorption. The weakest signal is associated with the $F110W$ broad-band filter because, contrary to $CH_4$-off, this passband covers a wide range of wavelengths much influenced by intense water vapor, ammonia, and methane absorptions (see next).

To constrain the atmospheric properties of J0855$-$0714, we compared its SED with state-of-the-art solar-metallicity planetary atmosphere models computed by \citet{morley14}. These models include the treatment of the water cloud opacity in cold atmospheres, which is found to have an impact on the emergent spectrum \citep[and references therein]{burrows04,sudarsky05}. The grid of models is available for effective temperatures ($T_{\rm eff}$) and surface gravities (log\,$g$) in the intervals 200--450 K and 3.0--5.0 (cm\,s$^{-2}$) with increments of 25--50 K and 0.5 dex, respectively. For all theoretical spectra used here, \citet{morley14} adopted $f_{\rm sed}$\,=\,5 and $h$\,=\,0.5, where $f_{\rm sed}$ is a parameter that describes the efficiency of sedimentation in the atmospheres and $h$ represents the fractional atmospheric area covered in cloud holes. The models also incorporate the salt and sulfide clouds (Na$_2$S, KCl, ZnS, MnS, Cr) and water-ice clouds. In Figure~\ref{allmodels} of the Appendix, various of these theoretical spectra are shown together with J0855$-$0714's SED. From the models, the near-infrared fluxes dramatically decline with decreasing temperature. At the coolest temperature of the computations (200 K), nearly all flux emerges in the mid-infrared. 

For an easy comparison between the observed and theoretical data, we computed the model photometric SED for each temperature and gravity using the filter passbands corresponding to the various observations of J0855$-$0714. The resulting theoretical photometric SEDs are also shown normalized to the target's $W2$-band emission in Figures~\ref{sed} and~\ref{allmodels}. \citet{morley14} spectra provide a reasonable description of the SED of J0855$-$0714. To find the best fit temperature and gravity, we minimized the following expression: 
\begin{equation}
\psi^2 = \sum_{i=1}^{n} \frac{[f_{\rm obs}(i)-f_{\rm mod}(i)]^2}{f_{\rm obs}^2(i)} \label{eq1}
\end{equation}
where $f$ stands for the observed and modeled fluxes for each wavelength $i$ in which there is a detection of J0855$-$0714 ($n=7$, $F110W$, $J$, $CH_4$-off, $W1$, $[3.6]$, $[4.5]$, and $W2$). The best fit (smallest value of $\psi^2$) is found for the theoretical computation with $T_{\rm eff}$\,=\,225 K and high gravity log\,$g$\,=\,5.0 (cm\,s$^{-2}$). In the minimization process, the most deviant point corresponds to $F110W$ (see next); models of $T_{\rm eff} > 225$ K and the $F110W$ observation come to differ by one and two orders of magnitude. Only the 225-K model provides a good match to this data point (see Figure~\ref{allmodels}). If $F110W$ is removed from Equation~\ref{eq1}, then the best fit solution is found for a temperature intermediate between 225 and 250 K and log\,$g$\,=\,5.0 (cm\,s$^{-2}$). Such low temperature agrees with the estimations made for J0855$-$0714 by \citet{luhman14a}, \citet{wright14}, \citet{faherty14}, \citet{beamin14}, \citet{kopytova14}, and \citet{leggett15}. In Figure~\ref{sed}, we plot an intermediate model spectrum (by averaging the 225-K and 250-K data) along with its computed photometry. The strongest absorption features are due to water, methane, and ammonia, with some tiny contribution from PH$_3$ in the mid-infrared at 4.3 $\mu$m. 

From Figure~\ref{sed}, it is seen that the $J$, $CH_4$-off and $[4.5]$ data of J0855$-$0714 are reproduced by the model within 1.5-$\sigma$ the quoted uncertainties (all data are normalized to the $W2$ emission of the target). Additionally, the observed flux upper limits at wavelengths $\ge$1.4 $\mu$m are consistent with the model predictions (this includes $H$, $K$, $W3$, and $W4$). However, there are discrepancies between the theory and the observations at certain wavelengths. It appears that the models envision a very strong methane absorption at 3--4 $\mu$m (see also Figure~\ref{allmodels}), which disagrees with the observations ($W1$ and $[3.6]$). At shorter wavelengths, neither $F110W$ nor the bluer filters $z$ and $Y$ are matched by the 225--250 K model. The synthetic spectra warmer than 225 K predict too much emission (or too less absorption) than expected at these wavelengths. $F110W$ includes the non-detection at $F105W$ \citep{schneider16}, and the two detections at $F125W$- and $J$-bands, the latter of which is reasonably reproduced. This hints at the models overpredicting the flux emission at wavelengths bluewards of $\sim$1.1 $\mu$m. These mismatches may indicate an incorrect or incomplete treatment of methane in current models. It appears that non-equilibrium carbon chemistry in cool atmospheres, possibly related to vertical flows of material, is required to explain the spectra of L and T dwarfs and the giant planets of the Solar System \citep[e.g.,][]{oppenheimer98,visscher11,currie14}.

\begin{figure}
%\resizebox{\hsize}{!}
\center{\includegraphics[scale=0.48]{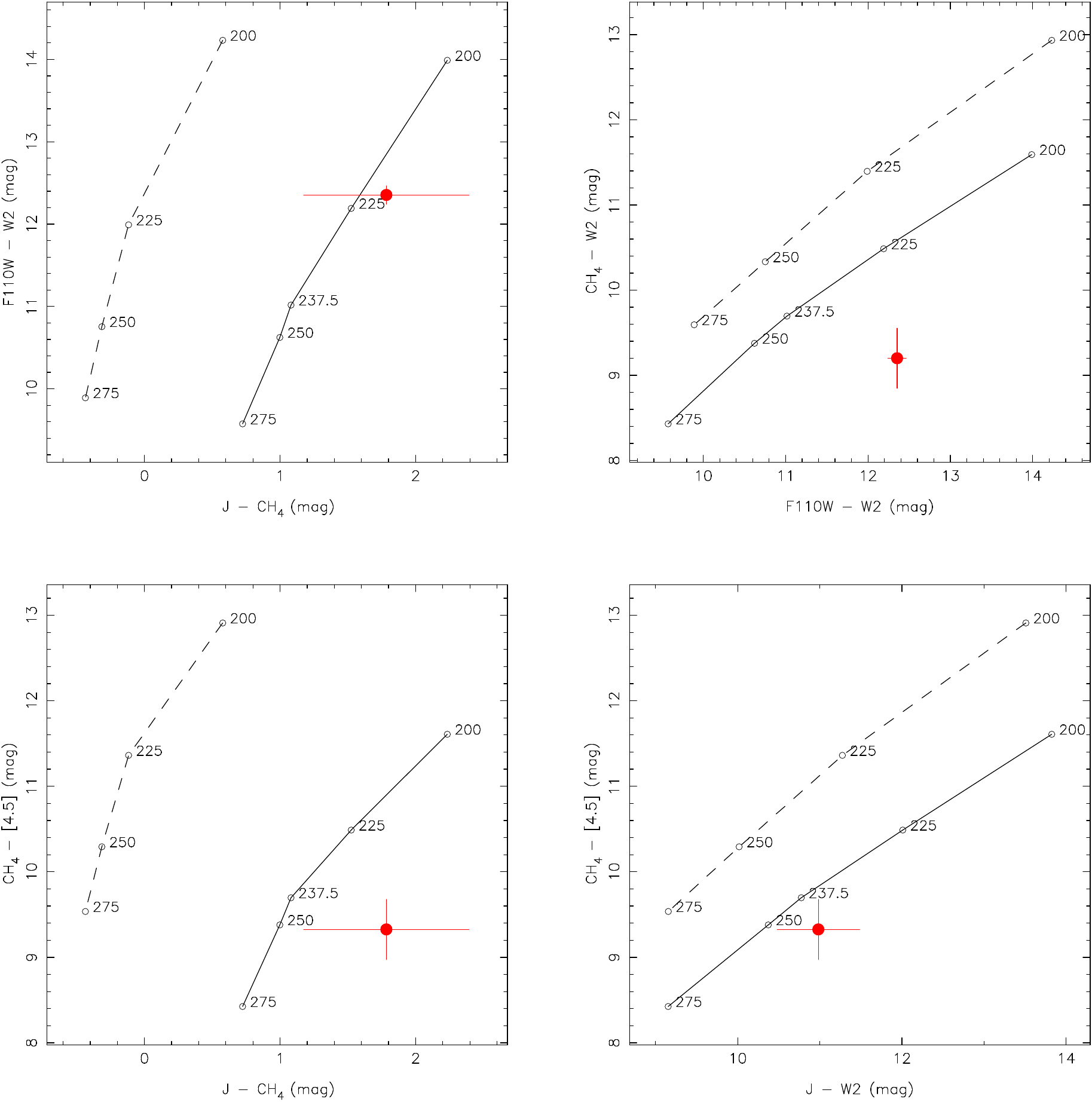}}
\caption{Color-color diagrams of J0855$-$0714 (red dot) including the synthetic indices (open circles) computed from the \citet{morley14} theoretical spectra. The high gravity colors (log\,$g$ = 5.0 cm\,s$^{-2}$) are joined by a solid line, and the dashed line stands for the low gravity indices (log\,$g$ = 4.0 cm\,s$^{-2}$). All synthetic colors are labeled with their corresponding $T_{\rm eff}$ (K). [$CH_4$ stands for the $CH_4$-off filter].
      \label{colors}}
\end{figure}

As illustrated in Figure~\ref{allmodels}, for a given temperature, the low gravity models envision more flux below $\sim$1 $\mu$m than the high gravity ones because of the weaker potassium absorption at low-pressure atmospheres. At the same time, $CH_4$-off becomes fainter at low gravities relative to the $J$-band fluxes. These two properties (as predicted by the \citealt{morley14} models) are not compatible with the SED of J0855$-$0714, for which we find that $CH_4$-off emission is stronger than the $J$-band fluxes in the units displayed in Figure~\ref{sed}. This supports the high-gravity nature of J0855$-$0714. 

Figure~\ref{colors} shows various color-color diagrams where the location of J0855$-$0714 is compared with the synthetic indices computed from the \citet{morley14} spectra. This Figure summarizes part of the discussion above. Note that according to these theoretical spectra, the impact of gravity is great at the low temperature regime, with notorious differences of about 1--1.5 mag (for a given temperature) in the near-infrared wavelengths. The inversion of the fluxes at $J$ and $CH_4$-off bands between the two gravities considered here is revealed by the blue (negative) and red (positive) $J-CH_4$-off colors displayed in the left panels of Figure~\ref{colors}. The two bottom panels of the Figure, which consider $J$, $CH_4$-off, $[4.5]$, and $W2$, are useful to discriminate $T_{\rm eff}$ and gravity for future J0855$-$0714-like discoveries. Nevertheless, the (large) differences observed between J0855$-$0714 and the theory in Figures~\ref{sed}, \ref{colors}, \ref{allmodels}, and~\ref{hr} (see next) indicate that some improvements in models of planetary atmosphers and evolution are necessary to better characterize this object.

By integrating the theoretical spectrum normalized to the $W2$-band of Figure~\ref{sed} from 0.6 through 50 $\mu$m and for the distance of 2.23\,$\pm$\,0.10 pc, we determined a bolometric luminosity of log\,$L/L_\odot$ = $-$8.57$\pm$0.06 dex for J0855$-$0714, where the uncertainty comes from the errors in $W2$ and distance. The flux excess below $\sim$1.1 $\mu$m given by the model does not have a significant impact on this determination because the great amount of the emission happens at $\sim$4.5 $\mu$m. The luminosity value reported here strongly depends on the models used. The true luminosity of J0855$-$0714 might be higher if its methane absorption at $\sim$3.5 $\mu$m is not as intense as the one predicted by the model.

\begin{figure}
%%\resizebox{\hsize}{!}
\center{\includegraphics[scale=0.5]{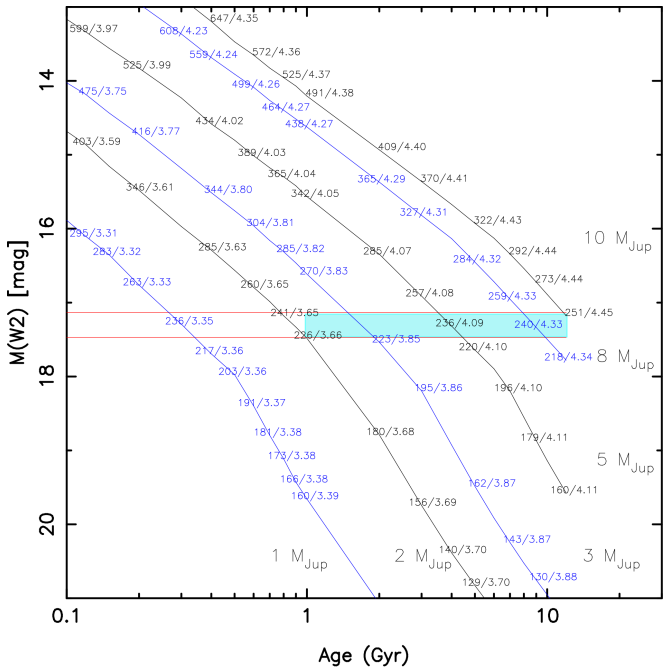}}
\caption{The evolution of the absolute $W2$ magnitude is shown for masses between 1 and 10 M$_{\rm Jup}$ \citep[COND models;][]{baraffe03}. Each mass track is labeled with the $T_{\rm eff}$/log\,$g$ pair values (K, cm\,s$^{-2}$) corresponding to the different ages tabulated in the models. J0855$-$0714's absolute $W2$ magnitude is bracketed by the two horizontal red lines (1 $\sigma$). The blue shaded area indicates the most likely position of J0855$-$0714 based on its high surface gravity [log\,$g \gtrapprox 3.5$ (cm\,s$^{-2}$)].
      \label{w2age}}
\end{figure}

\section{Mass and age}
We also compared the observed photometry and the derived $T_{\rm eff}$ and surface gravity with the models for cool brown dwarfs and extrasolar giant planets of \citet{baraffe03} to set constraints on the mass and age of J0855$-$0714. These models directly provide the magnitudes in the filters of interest (except for $CH_4$-off) by integrating over the theoretical spectra computed by \citet{allard01}. Figure~\ref{w2age} displays the absolute $W2$ magnitudes as a function of age and planetary mass. We also facilitate the $T_{\rm eff}$ and log\,$g$ theoretical values of the models in Figure~\ref{w2age} to aid the discussion below. Objects as cool as our target have their flux emission peak in this band. We relied on $W2$ to determine the most likely mass and age for J0855$-$0714. The high gravity obtained from Section~\ref{fit} indicates that this dwarf has a small radius compatible with an old age (young objects are undergoing a self-collapse process and have large radii, e.g., \citealt[and references therein]{lodieu15,kraus15}). However, the value of log\,$g$\,=\,5 (cm\,s$^{-2}$) exceeds by at least 0.5 dex those predicted for objects as cool as a few hundred K at the age of the Galaxy ($\sim$12 Gyr) by the evolutionary models of \citet{baraffe03}. We also found the same mismatch when using the evolutionary models of \citet{saumon08}. We overcame this discrepancy by qualitatively accepting that J0855$-$0714 has contracted sufficiently and that it likely has an age typical of the solar neighborhood \citep{holmberg09}. The age distribution of the solar vicinity peaks at 1--3 Gyr and rapidly delines toward higher ages \citep{nordstrom04}. For these ages, and according to the substellar cooling models shown in Figure~\ref{w2age}, high-gravity values may be defined by log\,$g \gtrapprox 3.5$ (cm\,s$^{-2}$) for low planetary-mass dwarfs.

The absolute $W2$ magnitude of J0855$-$0714 and its associated 1-$\sigma$ uncertainty [M($W2$) = 17.30\,$\pm$\,0.17 mag], which includes the errors in the observed photometry and the distance determination, is shown in Figure~\ref{w2age} with a band marked by two horizontal red lines. Interestingly, and contrary to the gravity parameter, the $T_{\rm eff}$'s inferred for the target dwarf from the evolutionary models agree with that obtained from the spectral fitting of Section~\ref{fit}. Note, however, that for all planetary masses and a given $W2$ magnitude the models give a similar $T_{\rm eff}$ for any of the ages illustrated in the Figure. This degeneracy is likely caused by the fact that the change in the planets' radius with age ($\ge$0.1 Gyr) is relatively small (less than 25\%); this is the size of the planets does not strongly depend on mass. To break this ambiguity, the surface gravity comes in handy. The blue shaded region depicted in Figure~\ref{w2age} considers high gravities in the interval log\,$g = 3.6 - 4.5$ (cm\,s$^{-2}$), from which we determined the mass of J0855$-$0714 to be 2--10 M$_{\rm Jup}$ for the age interval 1--12 Gyr. J0855$-$0714 would have a mass of $\approx$3 M$_{\rm Jup}$ at 2 Gyr, and a mass of $\approx$5 M$_{\rm Jup}$ at the age of the Solar System. Similar results are obtained from the {\sl Spitzer} $[4.5]$ data.

\begin{figure}
%\resizebox{\hsize}{!}
\center{\includegraphics[scale=0.5]{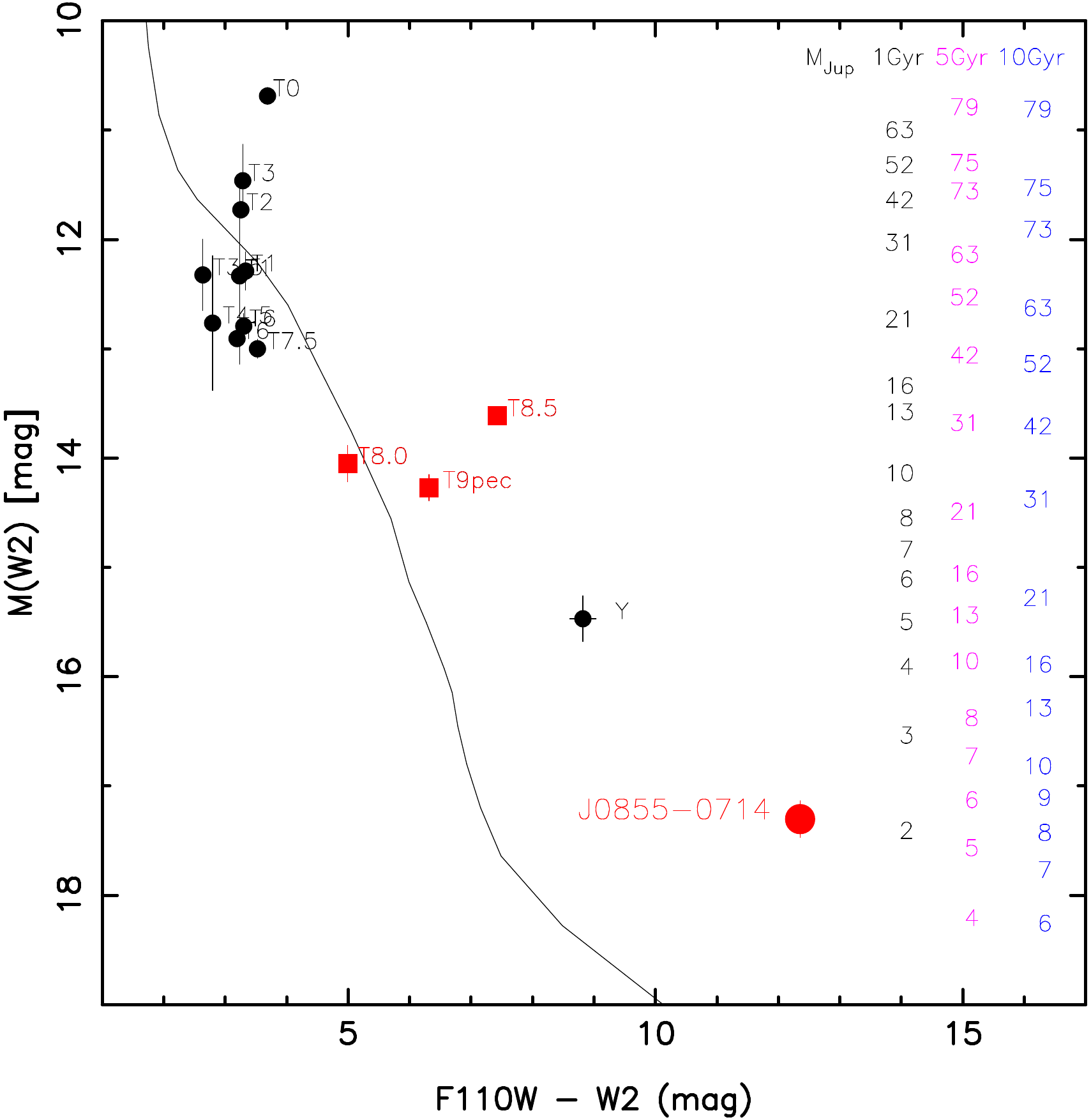}}
\caption{Color-magnitude diagram considering the $F110W$ and $W2$ filters. Our $F110W$ data are plotted as red symbols (J0855$-$0714 is labeled). Data from \citet{burgasser06} and \citet{luhman14c} are shown with black circles and labeled with their spectral types. [We assumed that $W2 \approx [4.5]$ for the Y-dwarf WD0806$-$661B \citep{luhman12a,luhman14c}. This assumption is based on the similarity of these two magnitudes in the case of J0855$-$0714]. The COND 5-Gyr evolutionary model of \citet{baraffe03} is depicted with a solid line. Masses in Jovian units predicted for the ages of 1, 5, and 10 Gyr are labeled to the right side.
      \label{hr}}
\end{figure}

Figure~\ref{hr} illustrates the faint tail of the sequence of ultra-cool dwarfs in the $W2$ versus $F110W - W2$ color-magnitude diagram. Data of T0--T7.5 dwarfs from \citet{burgasser06} and the Y-type WD0806$-$661B from  \citet{luhman12a,luhman14c} are shown together with J0855$-$0714 and three T8--T9 dwarfs for which we obtained $F110W$ photometry in a similar manner as for our target (see the Appendix). J0855$-$0714 stands out as the reddest and faintest source. The 5-Gyr isochrone of \citet{baraffe03} is also shown. Whereas it reasonably follows the trend pictured by the T dwarfs (despite the significant scatter in the observed data), the model is far from reproducing the extreme $F110W-W2$ color of cooler objects like our target, probably because of wrong predictions for the F110W filter, which is very sensitive to water, methane, and ammonia absorption (see Figure~\ref{sed}). While the massive near-infrared surveys like the VVV Survey \citep{minniti10} are discovering interesting cool nearby objects \citep[e.g.,][]{beamin13}, the diagram of Figure~\ref{hr} reveals that the mid-infrared observations turn critical for searching and characterizing the coldest planetary-mass objects. The near-infrared detections of J0855$-$0714 reported here are brighter than the limit of what could be detected by the Euclid\footnote{Euclid Definition Study Report (Red Book), ESA/SRE(2011)12.} mission, which is expected to reach $\sim$24.5 mag in $H$-band (3 $\sigma$, AB system) for the wide survey (between 15,000 and 20,000 deg$^2$), and two magnitudes fainter for the deep survey (40 deg$^2$). The combination of Euclid and NEOWISE \citep{schneider16} data will be a very effective way of selecting brown dwarf and free-floating planet candidates as cold as J0855$-$0714.

Based on the mass functions of young clusters \citep[e.g.,][]{bayo11,pena12,lodieu13}, we can estimate the number of J0855$-$0714-like objects populating the solar neighborhood by assuming that the mass distribution of stellar cluster members resembles that of the field population. In its power-law form (d$N$/d$M$ $\sim$ $M^{-\alpha}$), the mass function has a slope likely in the interval $\alpha$\,=\,0.4--1 for the low-mass stellar and substellar regimes \citep[see review by][]{luhman12b}. We would expect as many J0855$-$0714-like free-floating planets as 0.075--0.15 M$_\odot$ stars (spectral types $\sim$M5--L2) in the solar vicinity. Considering the possible values of the mass function exponent, this estimate can change by a factor of two. Their discovery is indeed challenging and will open a new window to the study of planetary atmospheres. Ground- and space-based mid-infrared instruments have the potential to play a significant role.

\section{Conclusions}
J0855$-$0714 is detected with a signal-to-noise ratio of $\approx$10 and $\approx$4 in the $F110W$ and $CH_4$-off bands of the WFC3 ({\sl HST}) and Hawk-I (VLT) instruments. This is the first detection in the $H$-band wavelengths, and the first $>$3-$\sigma$ ground-based detection at any wavelegnth. The comparison of the new photometry combined with the previously published data to state-of-the-art theoretical spectra computed for giant planets yields that J0855$-$0714 has a likely temperature of 225--250 K and a high surface gravity {[log\,$g \simeq 5$ (cm\,s$^{-2}$)]. J0855$-$0714 shows a red $J-CH_4$-off color of 1.78$\pm$0.61 mag in agreement with predictions of the high-gravity models and in marked contrast with the blue (negative) indices given by the low gravity synthetic spectra. However, the log\,$g \simeq 5$ (cm\,s$^{-2}$) obtained from the spectral fitting is not consistent with the predictions of evolutionary models of low-mass brown dwarfs and planets. For ages typical of the solar neighborhood (older than about 1 Gyr), the substellar evolutionary models envision a mass ranging from 2 to 10 M$_{\rm Jup}$  and gravities in the interval log\,$g$ = 3.5--4.5 (cm\,s$^{-2}$). If J0855$-$0714 has the age of the Sun, it would be a $\approx$5-M$_{\rm Jup}$ free-floating planetary-mass object. Based on the extrapolation of the stellar and substellar mass functions of young clusters, there could be as many J0855$-$0714-like sources in the solar neighborhood as low mass stars with spectral types M5--L2. At distances $<$7 pc, we estimate that 15--60 J0855$-$0714-like objects may be  present (based on the over 30 M5--L2 sources catalogued at a related distance from the Sun by the RECONS survey\footnote{www.recons.org}). Their discovery is indeed challenging.

\begin{acknowledgements}
Based on observations made with ESO Telescopes at the La Silla Paranal Observatory under program ID 094.C-0048(A) retrieved from the ESO Science Archive Facility. Based on observations made with the NASA/ESA {\sl Hubble Space Telescope}, obtained from the data archive at the Space Telescope Science Institute (STScI). STScI is operated by the Association of Universities for Research in Astronomy, Inc. under NASA contract NAS 5-26555. This research has made use of the Simbad and Vizier databases, operated at the Centre de Donn\'ees Astronomiques de Strasbourg (CDS), and of NASA's Astrophysics Data System Bibliographic Services (ADS). Current support for RECONS comes from the National Science Foundation. Our primary observing programs are carried out via the SMARTS Consortium, which operates four telescopes in the Chilean Andes under the auspices of National Optical Astronomy Observatory and the National Science Foundation. This research has been partly supported by the Spanish Ministry of Economy and Competitiveness (MINECO) under the grants AYA2014-54348-C3-2-R, AYA2015-69350-C3-2-P, and AYA2015-69350-C3-1. A$.$ B$.$ acknowledges financial support from the Proyecto Fondecyt de Iniciaci\'on 11140572. D$.$ M$.$ is supported by FONDECYT Regular No. 1130196, the BASAL CATA Center for Astrophysics and Associated Technologies  PFB-06, and the Ministry for the Economy, Development, and Tourism’s Programa Iniciativa Cient\'\i fica Milenio IC120009, awarded to the Millennium Institute of Astrophysics (MAS). K$.$ M$.$ acknowledges the support of the ESO-Government of Chile Joint Committee. J$.$ C$.$ B$.$ acknowledge support from CONICYT FONDO GEMINI - Programa de Astronom\'\i a del DRI, Folio 32130012. This work results within the collaboration of the COST Action TD 1308. 
\end{acknowledgements}

\bibliographystyle{aa} % style aa.bst
\bibliography{28662_zapatero_osorio} % your references Yourfile.bib

\appendix
\section{Additional material}
Table~\ref{log} provides the journal of the {\sl HST} and VLT observations of J0855$-$0714 publicly available.

Figure~\ref{allmodels} displays the comparison of J0855$-$0714's photometric SED with the theoretical spectra of \citet{morley14} computed for $f_{\rm sed} = 5$, $h = 0.5$ (half of the cloudy atmosphere is covered in holes), temperatures of 200--275 K, and two surface gravities (log\,$g$ = 4.0 and 5.0 cm\,s$^{-2}$). The best fit is given by the 225 K and log\,$g$\,=\,5.0 cm\,s$^{-2}$ model.

\begin{figure*}
%\resizebox{\hsize}{!}
\center{\includegraphics[scale=0.8]{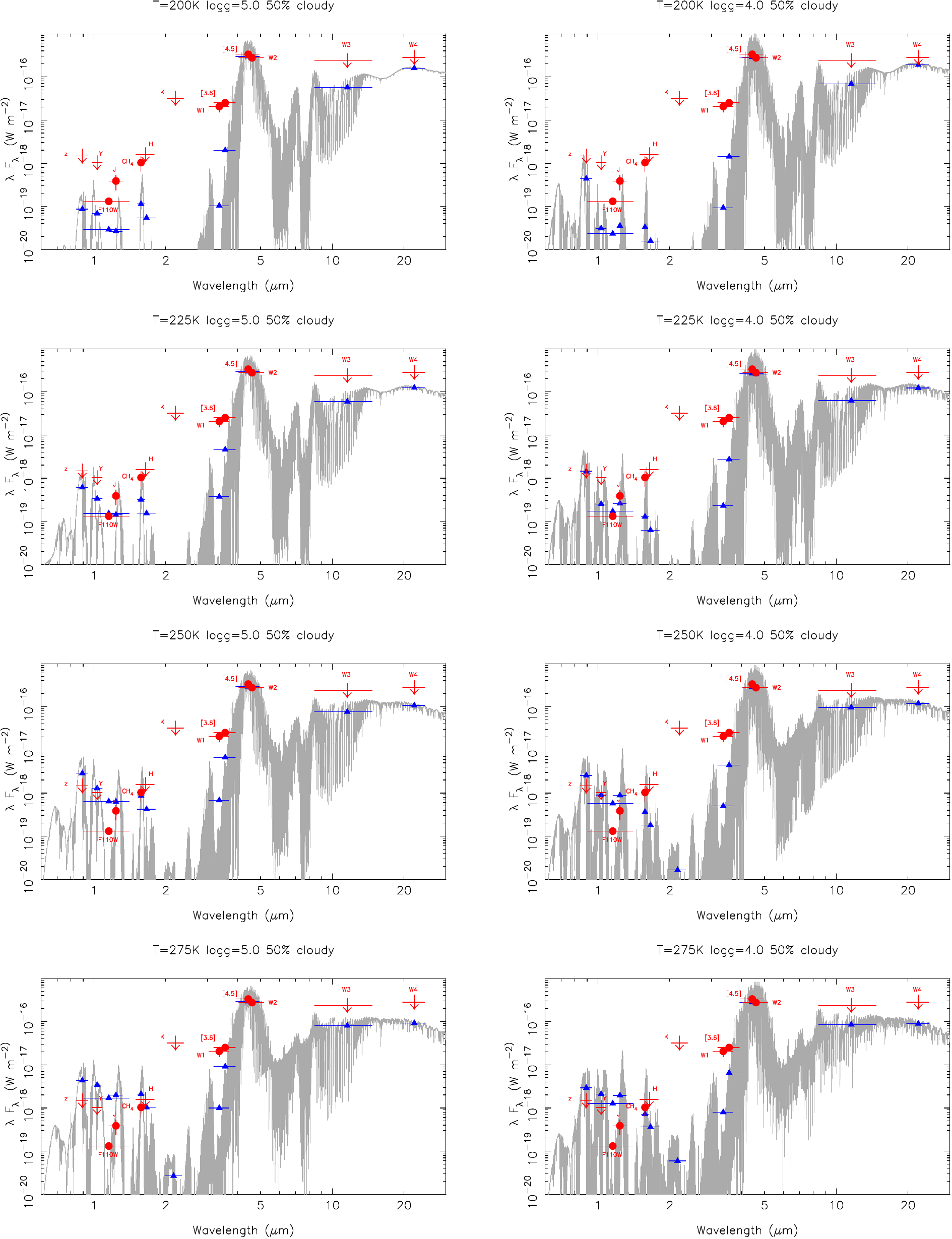}}
\caption{The spectral energy distribution of J0855$-$0714 (red symbols) is compared with various planetary atmosphere models (gray lines) of \citet{morley14}, all of which are computed for a 50\%~cloudy atmosphere. Arrows indicate upper limits. The observing filters are labeled. The models on the left and right columns are calculated for log\,$g$\,=\,5.0 and 4.0 (cm\,s$^{-2}$), respectively. The blue triangles represent the theoretical flux densities as integrated from the models using the corresponding filter passbands. All models are normalized to the emission of J0855$-$0714 at $W2$. 
      \label{allmodels}}
\end{figure*}

We searched for public data of $\ge$T8 and Y-type dwarfs observed with the $F110W$ filter and the WFC3 instrument on-board the {\sl HST} to complement the color-magnitude diagram of Figure~\ref{hr}. Several observations are available. In addition, we selected the ultra-cool dwarfs with trigonometric parallaxes available in the literature \citep{faherty12,tinney14} and with a clear detection in the WFC3 data. Our search resulted in three T8--T9 dwarfs listed in Table~\ref{otherT}, for which we measured their $F110W$ photometry in the same manner as for J0855$-$0714 (Section~\ref{obs}).

\begin{table}
\caption{Journal of {\sl HST} and VLT observations downloaded from their respective archives.}
\label{log}
\centering
\tiny
\tabcolsep=0.11cm
\begin{tabular}{c c c c c c}
\hline
Instrument  & Filter      &  Date    &   Exptime    &  UT range   & Prog.\ ID \cr
            &             &           &   (s)        &          &           \cr
\hline
HST WFC3   & $F110W$      & 2014 Nov 25 &  5417.61      & 18h 25m -- 20h 46m & 13802 \cr
HST WFC3   & $F110W$      & 2015 Mar 03 &  5417.61      & 17h 45m -- 21h 28m & 13802 \cr
HST WFC3   & $F110W$      & 2015 Apr 11 &  5417.61      & 01h 23m -- 06h 33m & 13802 \cr
VLT HAWK-I & $CH_{4}$-off & 2014 Dec 02 & 250$\times$20 & 06h 12m -- 08h 13m & 094.C-0048(A) \cr
VLT HAWK-I & $CH_{4}$-off & 2015 Jan 10 & 125$\times$20 & 03h 55m -- 04h 45m & 094.C-0048(A) \cr
VLT HAWK-I & $CH_{4}$-off & 2015 Jan 16 & 125$\times$20 & 04h 40m -- 05h 30m & 094.C-0048(A) \cr
VLT HAWK-I & $CH_{4}$-off & 2015 Jan 18 & 250$\times$20 & 03h 58m -- 05h 43m & 094.C-0048(A) \cr
VLT HAWK-I & $CH_{4}$-off & 2015 Jan 20 & 250$\times$20 & 03h 04m -- 04h 50m & 094.C-0048(A) \cr
VLT HAWK-I & $CH_{4}$-off & 2015 Jan 21 & 350$\times$20 & 03h 30m -- 04h 32m & 094.C-0048(A) \cr
\hline
\end{tabular}
\end{table}

%DATE-OBS= '2015-04-11'         / UT date of start of observation (yyyy-mm-dd)
%TIME-OBS= '04:27:11'           / UT time of start of observation (hh:mm:ss)
%EXPSTART=       57123.05798041 / exposure start time (Modified Julian Date)
%EXPEND  =       57123.27347745 / exposure end time (Modified Julian Date)
%EXPTIME =          5417.610354 / exposure duration (seconds)--calculated

%And the coordinates of WISE0855 are , 08:55:0.8.163 -07:14:39.34, see my email on 18 April.

\begin{table}
\caption{Additional $F110W$ photometry.}
\label{otherT}
\centering
\tiny
\tabcolsep=0.11cm
\begin{tabular}{llccccc}
\hline
Object                      & SpT  & $F110W$         &  $W2$          &  Date & Prog.\ ID & Exp$.$ time\\
                            &      & (mag)           & (mag)          &       &           &   (s) \\
\hline
ULAS\,J003402.77$-$005206.7 & T8.0 &  19.54$\pm$0.04 & 14.54$\pm$0.06 & 2009 Dec 27 & 11666 & 111.031 \\  
WISE\,J104245.23$-$384238.3 & T8.5 &  21.99$\pm$0.11 & 14.56$\pm$0.05 & 2013 Jun 10 & 12972 & 1211.739\\  
WISE\,J232519.53$-$410535.0 & T9pec&  20.43$\pm$0.05 & 14.11$\pm$0.04 & 2013 Jun 07 & 12972 & 1111.752  \\  
\hline
\end{tabular}
\end{table}

\end{document}